\documentclass[aps,pra,superscriptaddress,twocolumn]{revtex4}
\usepackage{amsmath}
\usepackage{amssymb}
\usepackage{graphicx}
\usepackage{dcolumn}
\usepackage{bm}

\begin{document}

\title{Strong parity effect of particle number in the interference fringes
of Bose-Einstein condensates released from a double-well potential}
\author{Hongwei Xiong}
\email{xionghongwei@wipm.ac.cn}
\affiliation{State Key Laboratory of Magnetic Resonance and Atomic and Molecular Physics,
Wuhan Institute of Physics and Mathematics, Chinese Academy of Sciences,
Wuhan 430071, P. R. China}
\affiliation{Center for Cold Atom Physics, Chinese Academy of Sciences, Wuhan 430071, P.
R. China}
\author{Shujuan Liu}
\affiliation{State Key Laboratory of Magnetic Resonance and Atomic and Molecular Physics,
Wuhan Institute of Physics and Mathematics, Chinese Academy of Sciences,
Wuhan 430071, P. R. China}
\affiliation{Center for Cold Atom Physics, Chinese Academy of Sciences, Wuhan 430071, P.
R. China}
\date{\today }

\begin{abstract}
We study the parity effect of the particle number in the interference
fringes of a Bose-Einstein condensate released from a double-well potential.
For a coherently splitting condensate in the double-well potential, with a
decoupled two-mode Bose-Hubbard model, there is well-known phase diffusion
because of interatomic interactions. After a specific holding time of the
double-well potential, the phase diffusion will make the interference
patterns in the density distribution depend strongly on the parity of the
total particle number by further overlapping two condensates. This parity
effect originates from the quantized relative phase about the total particle
number. The experimental scheme to observe this \textquotedblleft even-odd
\textquotedblright effect of the particle number is discussed.

%PACS: 03.75.Dg; 03.75.Nt; 03.75.Kk
\end{abstract}

\maketitle

%\email{sjliu@wipm.ac.cn}

\section{introduction}

For most classical systems with a large number of particles, adding or
decreasing one particle will not change the fundamental properties of the
system. For the quantum system of dilute Bose or Fermi gases in a harmonic
trap with a large number of particles, it is also widely believed that
adding or decreasing one particle will not change the thermodynamic and
dynamic properties of the system. For example, for a Bose-Einstein
condensate (BEC) described by the Gross-Pitaevskii equation within the
mean-field model, adding or decreasing one particle plays a negligible role
in the coupling parameter (which is proportional to the overall particle
number) for the order parameter. Recently, it is found theoretically that
for specific cold atomic system beyond the mean-field model, there is a
parity effect of the particle number which means that adding or decreasing
one particle would has different quantum effect. The parity effect has been
studied theoretically for the quantum decay of Josephson $\pi $ states \cite%
{Hat}, the Berry phase of two-species Bose-Einstein condensates (BECs) \cite%
{Chen}, the tunnel splitting in the Josephson model \cite{Lu}, the
macroscopic superpositions of phase states \cite{Piazza}, and more recently
in the mesoscopic quantum switching \cite{Shch}.

To our best knowledge, the parity effect of particle number has not been
observed experimentally for cold atoms, although the parity-dependent
tunneling splitting was observed for magnetic molecular clusters \cite%
{Hal,Loss,Delft,Wern}. It is still an open question to observe the
parity-dependent effect for cold atomic system. For example, it is very
difficult for an experiment to observe the parity-dependent effect in the
tunneling splitting, because the tunneling splitting is extremely small \cite%
{Shch}.

%It is necessary to consider theoretically different parity effect
%and other schemes to observe the parity effect.

The quantum interference between two condensates in a double-well potential
\cite%
{Andrew,Shin,Shin1,Schumm,Hoff,Hoff1,Ying,Chao,Wu,Hori,Burkov,Smith,Wang}
has renewed interest in the experimental and theoretical studies of the
macroscopic quantum coherence effect. In the last few years, there are
significant advances in the nonlinear self-trapping of weakly coupled
condensates \cite{Smerzi,Albiez}, long phase coherence time for two
separated condensates \cite{Jo} \textit{etc}. Stimulated by the remarkable
experimental advances of BECs in double-well potential, we find strong
parity-dependent interference fringes for ultracold bosonic gases released
from a double-well potential, based on the phase diffusion in two-mode
Bose-Hubbard model. For specific holding time of the double-well potential
to create the phase diffusion, we find that the center in the density
distribution is a dip for even particle number, while it is a peak for odd
particle number. This provides a way to display directly the quantized
characteristic of the particle number.

The paper is organized as follows. In Sec. II, we consider the phase
diffusion of a condensate in a double-well potential, with the decoupled
Bose-Hubbard Hamiltonian. The parity effect of the particle number is
discussed based on the coherence property. In Sec. III, we predict strong
parity effect of the particle number in the interference patterns of the
density distribution. In Sec. IV, the influence of the parity effect due to
asymmetry fluctuations of the double-well potential and particle number
squeezing is studied. In the last section, we give a brief summary and
discuss the application of Feshbach resonance to observe more clearly the
parity effect.

\section{phase diffusion and coherence property of a condensate in a
double-well potential}

For a condensate in a double-well potential with negligible tunneling, the
system can be described by the following decoupled Bose-Hubbard Hamiltonian.%
\begin{equation}
\widehat{H}=\sum_{i=1}^{2}\varepsilon _{i}\widehat{n}_{i}+\sum_{i=1}^{2}%
\frac{U}{2}\widehat{n}_{i}\left( \widehat{n}_{i}-1\right) .
\end{equation}%
Here $\widehat{n}_{i}=\widehat{a}_{i}^{\dag }\widehat{a}_{i}$ is the
particle number operator for the $i$th condensate. $\widehat{a}_{i}$ and $%
\widehat{a}_{i}^{\dag }$ are bosonic annihilation and creation operators for
the $i$th condensate. $\varepsilon _{1}-\varepsilon _{2}$ represents the
bias potential between two sites, while $U$ denotes collisional interaction
energy. The initial quantum state in this double-well potential is assumed as%
\begin{equation}
\left\vert \Psi \left( t_{h}=0\right) \right\rangle =\frac{1}{\sqrt{2^{N}N!}}%
\left( \widehat{a}_{1}^{\dag }+\widehat{a}_{2}^{\dag }\right) ^{N}\left\vert
0\right\rangle .  \label{initialstate}
\end{equation}%
The evolution of the quantum state is then
\begin{eqnarray}
\left\vert \Psi \left( t_{h}\right) \right\rangle &=&\frac{1}{\sqrt{2^{N}N!}}%
e^{-\frac{i}{\hbar }\widehat{H}t_{h}}\left( \widehat{a}_{1}^{\dag }+\widehat{%
a}_{2}^{\dag }\right) ^{N}\left\vert 0\right\rangle  \notag \\
&=&\sum\limits_{l=0}^{N}\sqrt{\frac{N!}{2^{N}l!\left( N-l\right) !}}  \notag
\\
&&e^{-\frac{iUt_{h}}{2\hbar }\left[ l\left( l-1\right) +\left( N-l\right)
\left( N-l-1\right) \right] }\left\vert l,N-l\right\rangle .
\label{quantumstate}
\end{eqnarray}%
We see that there is a phase diffusion \cite{Wright,Java,Castin,Imamoglu} in
the quantum state $\left\vert \Psi \left( t_{h}\right) \right\rangle $
because of the exponential factor. In addition, from the exponential factor
in $\left\vert \Psi \left( t_{h}\right) \right\rangle $, the time evolution
of $\left\vert \Psi \left( t_{h}\right) \right\rangle $ displays a periodic
behavior of $T=4\pi \hbar /U$. To give a clear discussion of the parity
effect, in getting Eq. (\ref{initialstate}), we consider the case of $%
\varepsilon _{1}=\varepsilon _{2}$.

The above evolution of $\left\vert \Psi \left( t_{h}\right) \right\rangle $
may be realized by first preparing two coherently separated condensates in a
double-well potential with large Josephson tunneling, so that the quantum
state can be described well by Eq. (\ref{initialstate}). The central barrier
is then increased non-adiabatically so that the tunneling between two
condensates can be omitted, while the quantum state of the system still
takes the form given by Eq. (\ref{initialstate}). This is similar to the
experimental studies of the collapse and revival of the quantum coherence of
the matter wave packet in an optical lattice \cite{Bloch}. This scheme was
also applied in Ref. \cite{Piazza} to consider the macroscopic
superpositions of phase states.

A direct way to display the coherence property of $\left\vert \Psi \left(
t_{h}\right) \right\rangle $ is to consider the time evolution of $K\left(
t_{h}\right) =\left\langle \Psi \left( t_{h}\right) \right\vert \left(
\widehat{a}_{1}^{\dag }\widehat{a}_{2}+\widehat{a}_{2}^{\dag }\widehat{a}%
_{1}\right) \left\vert \Psi \left( t_{h}\right) \right\rangle /N$, which
reflects the phase coherence and interference effect between two
condensates. From the expression (\ref{quantumstate}), we have%
\begin{eqnarray}
&&K_{12}\left( t_{h}\right) =\frac{\left\langle \Psi \left( t_{h}\right)
\right\vert \widehat{a}_{1}^{\dag }\widehat{a}_{2}\left\vert \Psi \left(
t_{h}\right) \right\rangle }{N}  \notag \\
&=&\sum\limits_{l=1}^{N}\frac{\left( N-1\right) !}{2^{N}\left( l-1\right)
!\left( N-l\right) !}e^{-\frac{iUt_{h}}{\hbar }\left( N-2l+1\right) }.
\end{eqnarray}%
The above equation can be rewritten as%
\begin{equation}
K_{12}\left( t_{h}\right) =\frac{e^{-\frac{iUt_{h}}{\hbar }\left( N-1\right)
}}{2^{N}}\sum\limits_{l=0}^{N-1}\frac{\left( N-1\right) !}{l!\left(
N-1-l\right) !}\left( e^{\frac{2iUt_{h}}{\hbar }}\right) ^{l}.
\end{equation}%
Based on the binomial theorem, we have%
\begin{equation}
K_{12}\left( t_{h}\right) =\frac{e^{-\frac{iUt_{h}}{\hbar }\left( N-1\right)
}}{2^{N}}\left( 1+e^{\frac{i2Ut_{h}}{\hbar }}\right) ^{N-1}.
\end{equation}%
In this situation, we have the following simple expression%
\begin{equation}
K_{12}\left( t_{h}\right) =\frac{\left( \cos \left( Ut_{h}/\hbar \right)
\right) ^{N-1}}{2}.
\end{equation}

From the above results, we have%
\begin{equation}
K\left( t_{h0}\right) =\left( \cos \left( 4\pi t_{h0}\right) \right) ^{N-1},
\label{Kth0}
\end{equation}%
where $t_{h0}=t_{h}/T$. For large $N$, by using the identity limit $%
\lim_{N\rightarrow \infty }\left( 1-x/N\right) ^{N}=e^{-x}$, the above
expression can be approximated very well as%
\begin{equation}
K\left( t_{h0}\right) \approx \sum_{n\geq 0}\left( -1\right) ^{n\left(
N-1\right) }e^{-\frac{N-1}{2}\left( 4\pi \left( t_{h0}-n/4\right) \right)
^{2}},
\end{equation}%
where $n$ is an integer. In this situation, the width of the peaks or dips
in $K\left( t_{h0}\right) $ is about $1/2\pi \sqrt{N}$. At $t_{h0}=0$, there
is ideal phase coherence between two condensates. As shown in Fig. 1(a) for $%
N=101$, with time increasing, the phase coherence disappears very rapidly
for $t_{h0}>1/2\pi \sqrt{N}$. With further time increasing, however, there
is a revival of the phase coherence. This behavior is understood from the
periodic behavior of the quantum state $\left\vert \Psi \left( t_{h0}\right)
\right\rangle $.

A unique parity effect of the particle number is shown by the expression (%
\ref{Kth0}) of $K\left( t_{h0}\right) $. For specific holding time such that
$\cos \left( 4\pi t_{h0}\right) =-1$, we see that $K\left( t_{h0}\right) =1$
for odd particle number, while $K\left( t_{h0}\right) =-1$ for even particle
number. This parity effect is shown further in Fig. 1(a) and Fig. 1(b) for $%
N=101$ and $N=100$, respectively.

\begin{figure}[tbp]
\includegraphics[width=0.75\linewidth,angle=270]{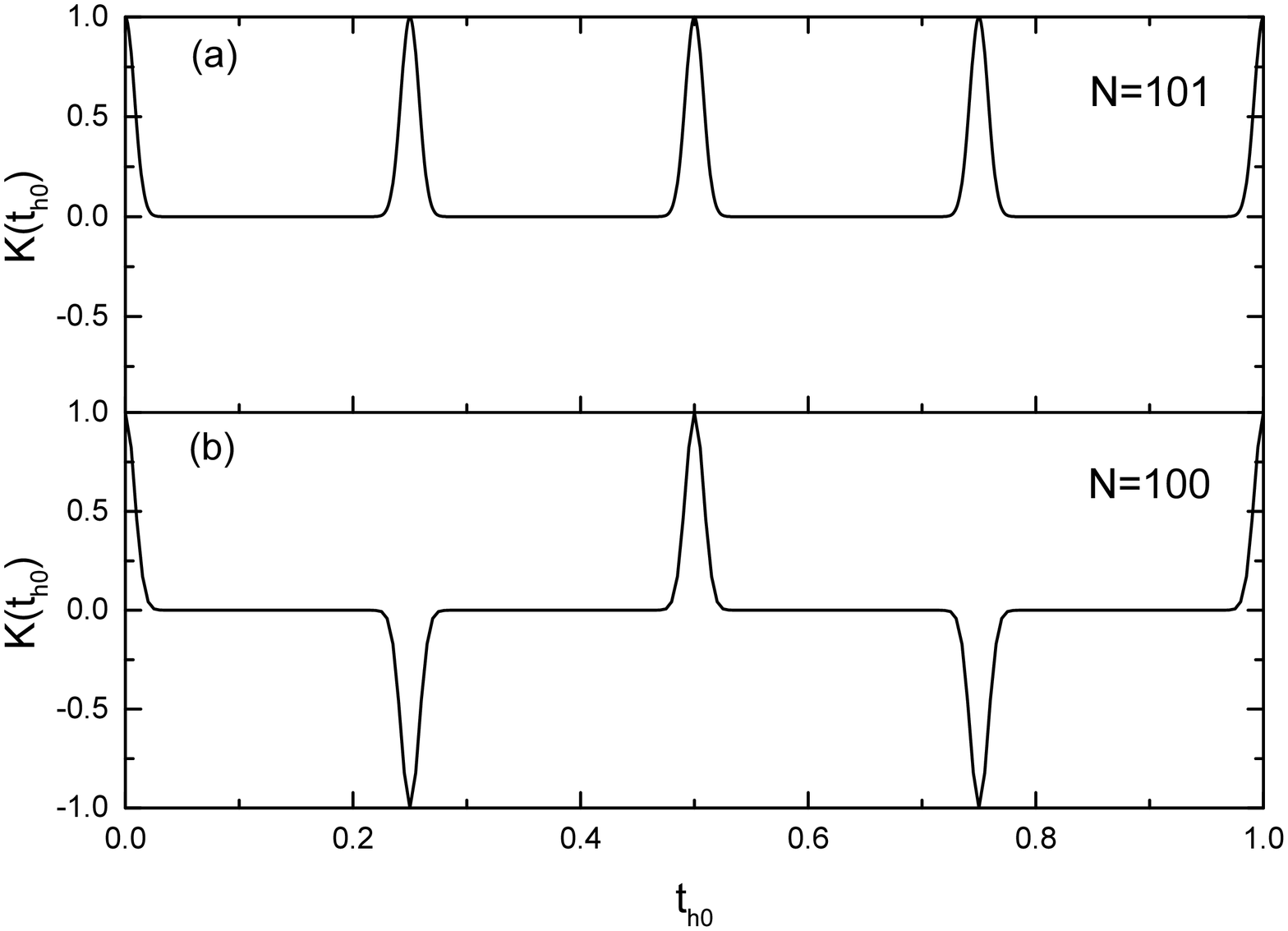}
\caption{Shown is the time evolution of the phase coherence factor $K\left(
t_{h0}\right) $. Here $t_{h0}$ is in unit of $T$ in the text. Besides the
periodic behavior of $K\left( t_{h0}\right) $, we see significantly
different behavior in $K\left( t_{h0}\right) $ for odd and even particle
number.}
\end{figure}

\section{parity effect of the particle number in the density distribution}

The above parity effect can be considered further by overlapping two
condensates after switching off the double-well potential. After a holding
time of $t_{h0}$, the quantum state is $\left\vert \Psi \left( t_{h0}\right)
\right\rangle $. Then, we calculate the density distribution after a time of
flight $t_{f}$ by switching off the double-well potential. In our
calculations, the initial wave functions of two condensates are assumed as $%
\varphi _{1}\left( x\right) =e^{-\left( x+d/2\right) ^{2}/2\sigma ^{2}}/\pi
^{1/4}\sigma ^{1/2}$ and $\varphi _{2}\left( x\right) =e^{-\left(
x-d/2\right) ^{2}/2\sigma ^{2}}/\pi ^{1/4}\sigma ^{1/2}$, respectively. In
this situation, the distance between two condensates is $d$. In the
following calculations, we will adopt the length unit $d$, energy unit $%
E_{d}=\hbar ^{2}/2md^{2}$ and time unit $T_{d}=\hbar /E_{d}$. After a time
of flight $t_{f}=t_{f0}T_{d}$, the density distribution of the system is
given by%
\begin{eqnarray}
n\left( x,t_{h0},t_{f0}\right) =\frac{N}{2}\left[ \left\vert \phi _{1}\left(
x,t_{f0}\right) \right\vert ^{2}+\left\vert \phi _{2}\left( x,t_{f0}\right)
\right\vert ^{2}+\right.  \notag \\
\left. 2\times \mathrm{Re}\left( \left( \cos \left( 4\pi t_{h0}\right)
\right) ^{N-1}\phi _{1}^{\ast }\left( x,t_{f0}\right) \phi _{2}\left(
x,t_{f0}\right) \right) \right] .
\end{eqnarray}%
For $N=100$ and $N=101$, in Figs. 2(a)-(f), we give the density distribution
for fixed time of flight $t_{f0}=1$ and different holding time $t_{h0}$. In
our calculations, $\sigma =0.1$. The factor $\left( \cos \left( 4\pi
t_{h0}\right) \right) ^{N-1}$ in the interference term of the above equation
gives several unique results:

\begin{figure}[tbp]
\includegraphics[width=0.75\linewidth,angle=270]{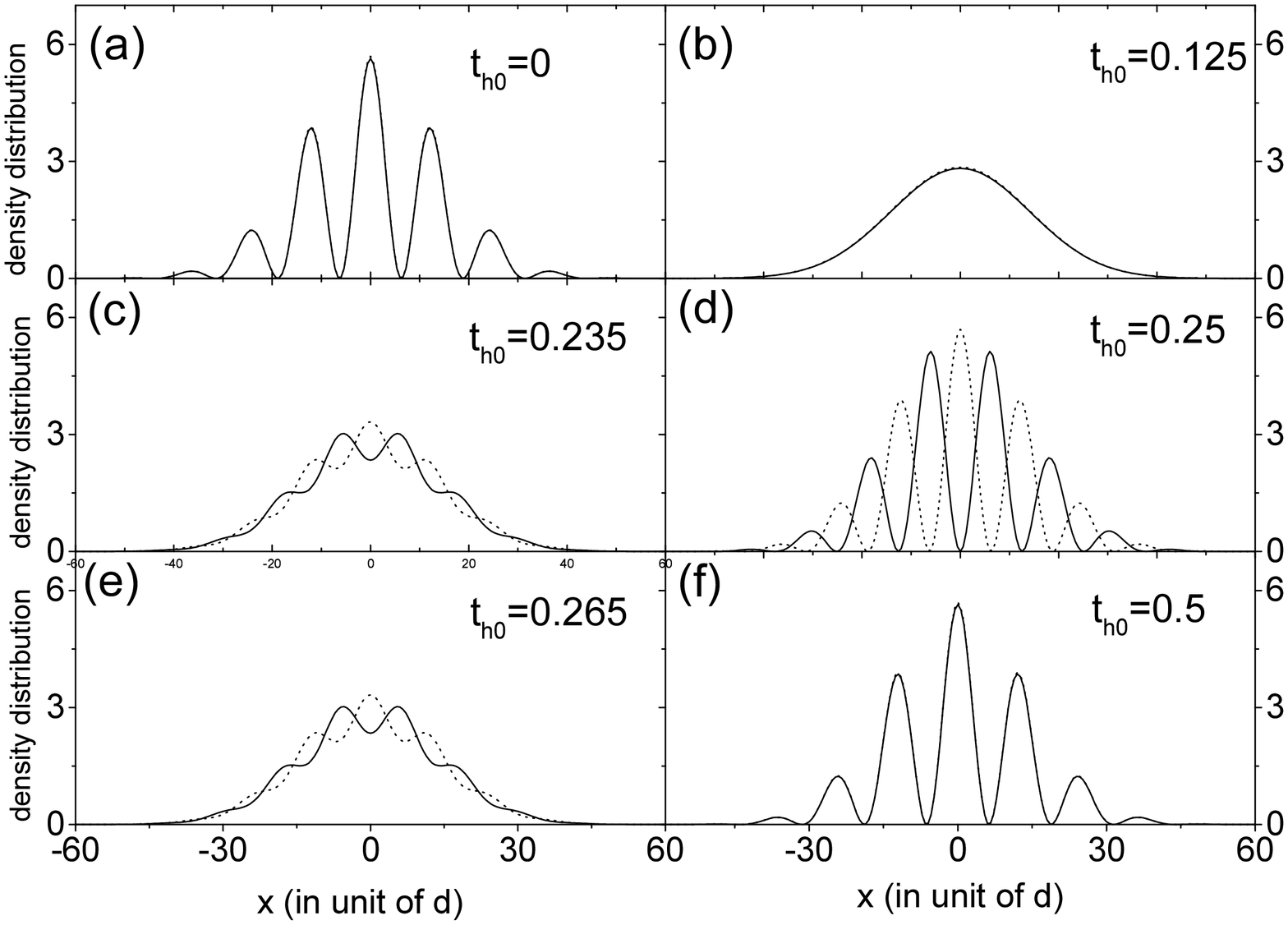}
\caption{(Color online) Shown is the density distribution for different
holding time $t_{h0}$ and fixed time of flight $t_{f0}=1$. The solid
(dotted) line is the density distribution of the particle number $N=100$ ($%
N=101)$. Clear parity effect of the particle number in the interference
patterns of the density distribution is shown in Figs. 2(c)-2(e).}
\end{figure}

(i) There are collapse and revival of the interference fringes in the
density distribution.

(ii) At $t_{h0}=j/2-0.25$ ($j$ is a natural number), the density
distribution depends strongly on the parity of the total particle number. As
shown in Fig. 2(d), the center of the density distribution (solid line) is a
dip for even particle number, while the center of the density distribution
(dotted line) is a peak for odd particle number. This can be understood
further by noticing that $n=N\left\vert \phi _{1}-\phi _{2}\right\vert
^{2}/2 $ for even particle number, while $n=N\left\vert \phi _{1}+\phi
_{2}\right\vert ^{2}/2$ for odd particle number. For both situations, the
density distribution is symmetric, which means a clear parity effect about
the total particle number \cite{comment-smerzi}.

At $t_{h0}=j/2-0.25$, the factor $\left[ \cos \left( 4\pi t_{h0}\right) %
\right] ^{N-1}$ in the interference term can be rewritten as $e^{i\pi \left(
N-1\right) }$. In this situation, the phase diffusion of the left and right
condensates is zero. However, there is a relative phase between two
condensates. This relative phase is quantized about the total particle
number. For even particle number, the relative phase is $\pi $, while the
relative phase is $2\pi $ for odd particle number.

Before the atomic cloud is imaged on a CCD camera, we do not know the parity
of the particle number. At $t_{h0}=j/2-0.25$, the parity effect would lead
to a random behavior in the density distribution. Averaging the density
distribution in different experiments, there would be no interference
patterns. However, in a single experiment, there will be clear interference
patterns, and there is $50\%$ probability to observe a peak (dip) in the
center of the density distribution. This is significantly different from the
holding time of $t_{h0}=j/2-0.5$, where one would always observe the same
interference patterns in the density distribution.

These analyses at $t_{h0}=j/2-0.25$ can be considered further by the
following density matrix%
\begin{equation}
\rho =P\left( N\right) \left\vert N\right\rangle \left\langle N\right\vert .
\end{equation}%
Here $P\left( N\right) $ is the probability to have $N$ atoms in the
condensate. It can be assumed as a Gaussian distribution about the average
particle number $\overline{N}$. The above expression can be rewritten as%
\begin{equation}
\rho =\sum\limits_{N=even}P\left( N\right) \left\vert N\right\rangle
\left\langle N\right\vert +\sum\limits_{N=odd}P\left( N\right) \left\vert
N\right\rangle \left\langle N\right\vert .
\end{equation}%
The ensemble average of the density distribution is then%
\begin{eqnarray}
&&\overline{n}\left( x,t_{h0},t_{f0}\right) =\mathrm{Tr}\left[ \rho \widehat{%
\Psi }^{\dag }\widehat{\Psi }\right]  \notag \\
&=&\frac{\overline{N}}{2}\left[ \left\vert \phi _{1}\left( x,t_{f0}\right)
\right\vert ^{2}+\left\vert \phi _{2}\left( x,t_{f0}\right) \right\vert ^{2}%
\right] .
\end{eqnarray}%
It is clear that there is no interference term in the ensemble average of
the density distribution. In a single-shot density distribution, however,
there would be interference fringes with random relative phase $\pi $ or $%
2\pi $.

\section{influence of the parity effect due to asymmetry of the double-well
potential and particle number squeezing}

As shown in preceding section, the density distribution depends strongly on
the parity of the overall particle number $N$. It is a natural problem
whether slight asymmetry fluctuations of the double-well potential will
destroy the parity effect. Generally speaking, the asymmetry fluctuations of
the double-well potential will lead to a random relative phase and
fluctuations of the average particle number in two sites. To consider both
effects, we study the following more general initial state%
\begin{equation}
\left\vert \Psi \left( t_{h}=0\right) \right\rangle =\frac{1}{\sqrt{2^{N}N!}}%
\left( \alpha \widehat{a}_{1}^{\dag }+\beta \widehat{a}_{2}^{\dag }\right)
^{N}\left\vert 0\right\rangle .
\end{equation}%
The normalization condition requests that $\left\vert \alpha \right\vert
^{2}+\left\vert \beta \right\vert ^{2}=2$. At a holding time $t_{h0}$, the
density distribution after a time of flight $t_{f0}$ is given by%
\begin{eqnarray}
&&n\left( x,t_{h0},t_{f0}\right) =\frac{N}{2}\left\{ \left\vert \alpha
\right\vert ^{2}\left\vert \phi _{1}\left( x,t_{f0}\right) \right\vert
^{2}+\left\vert \beta \right\vert ^{2}\left\vert \phi _{2}\left(
x,t_{f0}\right) \right\vert ^{2}\right.  \notag \\
&&+2\times \text{\textrm{Re}}\left[ \alpha ^{\ast }\beta \left( \frac{%
\left\vert \alpha \right\vert ^{2}e^{i4\pi t_{h0}}+\left\vert \beta
\right\vert ^{2}e^{-i4\pi t_{h0}}}{2}\right) ^{N-1}\right.  \notag \\
&&\left. \left. \phi _{1}^{\ast }\left( x,t_{f0}\right) \phi _{2}\left(
x,t_{f0}\right) \right] \right\} .
\end{eqnarray}

At the holding time $t_{h0}=j/2-0.25$, the density distribution is then%
\begin{eqnarray}
&&n\left( x,t_{h0},t_{f0}\right) =\frac{N}{2}\left\{ \left\vert \alpha
\right\vert ^{2}\left\vert \phi _{1}\left( x,t_{f0}\right) \right\vert
^{2}+\left\vert \beta \right\vert ^{2}\left\vert \phi _{2}\left(
x,t_{f0}\right) \right\vert ^{2}\right.  \notag \\
&&\left. +2\left( -1\right) ^{N-1}\text{\textrm{Re}}\left[ \alpha ^{\ast
}\beta \phi _{1}^{\ast }\left( x,t_{f0}\right) \phi _{2}\left(
x,t_{f0}\right) \right] \right\} .
\end{eqnarray}%
We see that there is still strong parity effect about the total particle
number when the asymmetry of the average particle numbers in two sites is
considered. For $t_{h0}=j/2-0.25$, $n=N\left\vert \alpha \phi _{1}-\beta
\phi _{2}\right\vert ^{2}/2$ for even particle number, while $n=N\left\vert
\alpha \phi _{1}+\beta \phi _{2}\right\vert ^{2}/2$ for odd particle number.

The factor $\alpha ^{\ast }\beta $ includes a random relative phase if the
fluctuations of the bias potential $\varepsilon _{1}-\varepsilon _{2}$ are
considered. The bias potential $\varepsilon _{1}-\varepsilon _{2}$ between
two sites will gives a factor $e^{-i\left( \varepsilon _{1}-\varepsilon
_{2}\right) t_{h}/\hbar }$. Experimentally, to observe the parity effect,
this requests that the fluctuations of the bias potential should satisfy $%
\left\vert \Delta \varepsilon \right\vert t_{h}/\hbar <<\pi $. This request
can be satisfied in the present experimental technique \cite{Schumm,Jo,Jo1}.
These analyses show that slight asymmetry fluctuations of the double-well
potential would not lead to serious problem in observing the parity effect.

In the non-adiabatical increasing of the central barrier so that the
tunneling between two condensates can be omitted, it is possible that there
would be a particle number squeezing. In this situation, we consider the
following quantum state%
\begin{equation}
\left\vert \Psi \left( t_{h}=0\right) \right\rangle
=\sum\limits_{l=0}^{N}\gamma e^{-\left( l-N/2\right) ^{2}s^{2}/N}\left\vert
l,N-l\right\rangle .
\end{equation}%
Here $s$ is the squeezing factor, and $\gamma $ is a normalization factor.
For this quantum state, we have%
\begin{eqnarray}
&&K_{12}\left( t_{h}\right) =\frac{\gamma ^{2}}{N}\sum\limits_{l=1}^{N}\sqrt{%
l\left( N-l+1\right) }  \notag \\
&&e^{-\left[ \left( l-N/2\right) ^{2}+\left( l-N/2-1\right) ^{2}\right]
s^{2}/N}e^{-\frac{iUt_{h}}{\hbar }\left( N-2l+1\right) }.
\end{eqnarray}%
At $t_{h0}=j/2-0.25$, we have%
\begin{eqnarray}
K_{12} &=&\frac{\gamma ^{2}}{N}\sum\limits_{l=1}^{N}\sqrt{l\left(
N-l+1\right) }  \notag \\
&&e^{-\left[ \left( l-N/2\right) ^{2}+\left( l-N/2-1\right) ^{2}\right]
s^{2}/N}e^{i\pi \left( N+1\right) }.
\end{eqnarray}%
At this holding time, we can also get%
\begin{eqnarray}
K_{21} &=&\frac{\left\langle \Psi \right\vert \widehat{a}_{2}^{\dag }%
\widehat{a}_{1}\left\vert \Psi \right\rangle }{N}  \notag \\
&=&\frac{\gamma ^{2}}{N}\sum\limits_{l=0}^{N-1}\sqrt{\left( l+1\right)
\left( N-l\right) }  \notag \\
&&e^{-\left[ \left( l-N/2\right) ^{2}+\left( l-N/2+1\right) ^{2}\right]
s^{2}/N}e^{i\pi \left( N+1\right) }.
\end{eqnarray}%
We see that at $t_{h0}=j/2-0.25$, even particle number corresponds to a
relative phase of $\pi $, while odd particle number corresponds to a
relative phase of $2\pi $. This means that the squeezing factor will not
destroy the parity effect in the interference fringes of the density
distribution. Of course, there is a request that the quantum state should
not be a Fork state in the extreme squeezing situation.

\section{summary and discussion}

In summary, a strong parity effect of the total particle number in the
density distribution of a condensate released from a double-well potential
is predicted in this work. The asymmetry fluctuations of the double-well
potential and particle number squeezing are discussed, and our studies show
that they would not lead to serious problem in observing the parity effect.
The parity effect is due to the quantized relative phase about the total
particle number for specific holding time of the double-well potential.
Another example of the strong parity effect about the total particle number
is the exchange effect for a complex of $N$ spin $1/2$ fermions \cite{Kohn}.
The parity effect of the complex lies in that it is a boson for even $N$,
while it is a fermion for odd $N$.

In this work, similarly to the phase diffusion considered in Ref. \cite{Java}%
, we ignore the phase diffusion in the increasing of the central barrier by
assuming that the central barrier is increased fast enough, and we also
ignore the phase diffusion in the time of flight. During the time of flight,
the interaction energy per particle will decrease rapidly, and thus the
phase diffusion may be ignored. In Ref. \cite{Piazza}, the phase diffusion
in the time of flight is also ignored. Nevertheless, overcoming these two
phase diffusions would contribute to more clear observation of the parity
effect. One way to overcome these two phase diffusions would be the
application of Feshbach resonance which has been used in atom interferometry
\cite{Widera,Fattori}. The scheme with the application of Feshbach resonance
would be: (i) One first prepares two coherently separated condensates in a
double-well potential. (ii) After turning the s-wave scattering length
almost to zero via a magnetic-field Feshbach resonance, the central barrier
is increased adiabatically so that the tunneling between two sites can be
omitted. (iii) Holding the double-well potential, one adiabatically
increases and then decreases the s-wave scattering length almost to zero so
that the condition $\int_{0}^{t_{h}}\frac{U\left( t\right) }{\hbar }dt=\pi $
is satisfied. (iv) Switching off the double-well potential, the ideal gas is
then imaged after a time of flight to display the parity effect.

\begin{acknowledgments}
We acknowledge useful discussions with Prof. Y. Wu, Prof. B. Wu, Prof. B. L.
Lv and Prof. L. You. This work was supported by NSFC under Grant Nos.
10875165, 10804123, 10634060, and NKBRSF of China under Grant No.
2006CB921406.
\end{acknowledgments}


\begin{thebibliography}{99}
\bibitem{Hat} N. Hatakenaka, Phys. Rev. Lett. \textbf{81}, 3753 (1998).

\bibitem{Chen} Z. D. Chen, J. Q. Liang, S. Q. Shen, and W. F. Xie, Phys.
Rev. A \textbf{69}, 023611 (2004).

\bibitem{Lu} R. L\H{u}, M. Zhang, J. L. Zhu, and L. You, Phys. Rev. A
\textbf{78}, 011605(R) (2008).

\bibitem{Piazza} F. Piazza, L. Pezz\'{e}, and A. Smerzi, Phys. Rev. A
\textbf{78}, 051601(R) (2008).

\bibitem{Shch} V. S. Shchesnovich, Preprint arXiv: 0905.1708v2 (2009).

\bibitem{Hal} F. D. M. Haldane, Phys. Rev. Lett. \textbf{61}, 1029 (1988).

\bibitem{Loss} D. Loss, D. P. DiVincenzo, and G. Grinstein, Phys. Rev. Lett.
\textbf{69}, 3232 (1992).

\bibitem{Delft} J. von Delft and C. L. Henley, Phys. Rev. Lett. \textbf{69},
3236 (1992).

\bibitem{Wern} W. Wernsdorfer and R. Sessoli, Science \textbf{284}, 133
(1999).

\bibitem{Andrew} M. R. Andrews, C. G. Townsend, H. -J. Miesner, D. S.
Durfee, D. M. Kurn, and W. Ketterle, Science \textbf{275}, 637 (1997).

\bibitem{Shin} Y. Shin, M. Saba, T. A. Pasquini, W. Ketterle, D. E.
Pritchard, and A. E. Leanhardt, Phys. Rev. Lett. \textbf{92}, 050405 (2004).

\bibitem{Shin1} Y. Shin, G. B. Jo, M. Saba, T. A. Pasquini, W. Ketterle, and
D. E. Pritchard, Phys. Rev. Lett. \textbf{95}, 170402 (2005).

\bibitem{Schumm} T. Schumm, S. Hofferberth, L. M. Andersson, S. Wildermuth,
S. Groth, I. Bar-Joseph, J. Schmiedmayer, and P. Kruger, Nature Physics
\textbf{1}, 57 (2005).

\bibitem{Hoff} S. Hofferberth, I. Lesanovsky, B. Fischer, T. Schumm, and J.
Schmiedmayer, Nature \textbf{449}, 324 (2007).

\bibitem{Hoff1} S. Hofferberth, I. Lesanovsky, T. Schumm, A. Imambekov, V.
Gritsev, E. Demler, and J. Schmiedmayer, Nature Physics \textbf{4}, 489
(2008).

\bibitem{Ying} Y. Wu and X. X. Yang, Phys. Rev. A \textbf{68}, 013608 (2003).

\bibitem{Chao} C. H. Lee, Phys. Rev. Lett. \textbf{97}, 150402 (2006).

\bibitem{Wu} B. Wu and J. Liu, Phys. Rev. Lett. \textbf{96}, 020405 (2006).

\bibitem{Hori} M. Horikoshi and K. Nakagawa, Phys. Rev. Lett. \textbf{99},
180401 (2007).

\bibitem{Burkov} A. A. Burkov, M. D. Lukin, and E. Demler, Phys. Rev. Lett.
\textbf{98}, 200404 (2007).

\bibitem{Smith} K. Smith-Mannschott, M. Chuchem, M. Hiller, T. Kottos, and
D. Cohen, Phys. Rev. Lett. \textbf{102}, 230401 (2009).

\bibitem{Wang} J. Wang and J. B. Gong, Phys. Rev. Lett. \textbf{102}, 244102
(2009).

\bibitem{Smerzi} A. Smerzi, S. Fantoni, S. Giovanazzi, and S. R. Shenoy,
Phys. Rev. Lett. \textbf{79}, 4950 (1997).

\bibitem{Albiez} M. Albiez, R. Gati, J. F\H{o}lling, S. Hunsmann, M.
Cristiani, and M. K. Oberthaler, Phys. Rev. Lett. \textbf{95}, 010402 (2005).

\bibitem{Jo} G. B. Jo, Y. Shin, S. Will, T. A. Pasquini, M. Saba, W.
Ketterle, D. E. Pritchard, M. Vengalattore, and M. Prentiss, Phys. Rev.
Lett. \textbf{98}, 030407 (2007).

\bibitem{Wright} E. M. Wright, D. F. Walls, and J. C. Garrison, Phys. Rev.
Lett. \textbf{77}, 2158 (1996).

\bibitem{Java} J. Javanainen and M. Wilkens, Phys. Rev. Lett. \textbf{78},
4675 (1997).

\bibitem{Castin} Y. Castin and J. Dalibard, Phys. Rev. A \textbf{55}, 4330
(1997).

\bibitem{Imamoglu} A. Imamoglu, M. Lewenstein, and L. You, Phys. Rev. Lett.
\textbf{78}, 2511 (1997).

\bibitem{Bloch} M. Greiner, O. Mandel, T. W. Hansch, and I. Bloch, Nature
\textbf{419}, 51 (2002).

\bibitem{comment-smerzi} In preparing this manuscript, we noticed the work
in Ref. \cite{Piazza}, where decoupled Bose-Hubbard Hamiltonian is also used
to predict a parity effect of the total particle number. In the density
distribution shown in Fig. 2(d) of our paper, the holding time is $%
t_{h0}=0.25$. In Ref. \cite{Piazza}, however, the holding time $t_{h0}=0.125$
is studied for the double-well situation, which predicts the macroscopic
superpositions of phase states.

\bibitem{Jo1} G. B. Jo, J.-H. Choi, C. A. Christensen, T. A. Pasquini, Y.-R.
Lee, W. Ketterle, and D. E. Pritchard, Phys. Rev. Lett. \textbf{98}, 180401
(2007).

\bibitem{Kohn} W. Kohn and D. Sherrington, Rev. Mod. Phys. \textbf{42}, 1
(1970).

\bibitem{Widera} A. Widera, S. Trotzky, P. Cheinet, S. F\H{o}lling, F.
Gerbier, I. Bloch, V. Gritsev, M. D. Lukin, and E. Demler, Phys. Rev. Lett.
\textbf{100}, 140401 (2008).

\bibitem{Fattori} M. Fattori, C. D'Errico, G. Roati, M. Zaccanti, M.
Jona-Lasinio, M. Modugno, M. Inguscio, and G. Modugno, Phys. Rev. Lett.
\textbf{100}, 080405 (2008).
\end{thebibliography}
\end{document}